\newcommand{\changed}[1]{#1}

\documentclass[twocolumn,amsthm,secthm]{autart}    

\usepackage{graphicx}          

\usepackage{subfig}
\usepackage{amsmath}
\usepackage{amsfonts}
\usepackage{amssymb}  
\usepackage{bm}
\usepackage{siunitx}
\usepackage{xspace}
\usepackage{xfrac}
\usepackage{empheq}
\usepackage{enumitem}
\usepackage{xcolor}
\usepackage[round]{natbib}

\begin{document}

\begin{frontmatter}

\title{Integral State-Feedback Control of Linear Time-Varying Systems: A Performance Preserving Approach} 

\thanks{The financial support by the Christian Doppler Research Association, the Austrian Federal Ministry for Digital and Economic Affairs and the National Foundation for Research, Technology and Development is gratefully acknowledged.
This work was partially supported by the Graz University of Technology LEAD project ``Dependable Internet of Things in Adverse Environments''.
}

\author[IRTCD]{Richard Seeber}\ead{richard.seeber@tugraz.at},
\author[IRT]{Markus Tranninger}\ead{markus.tranninger@tugraz.at}

\address[IRTCD]{Christian Doppler Laboratory for Model Based Control of Complex Test Bed Systems, Institute of Automation and Control, Graz University of Technology, Graz, Austria}
\address[IRT]{Institute of Automation and Control, Graz University of Technology, Graz, Austria}


\begin{abstract}                          An integral extension of state-feedback controllers for linear time-varying plants is proposed, which preserves performance of the nominal controller in the unperturbed case.
Similar to time-invariant state feedback with integral action, the controller achieves complete rejection of disturbances whose effective action on the plant is constant with respect to the control input.
Moreover, bounded-input bounded-state stability with respect to arbitrary disturbances is shown.
A modification for preventing controller windup as well as tuning guidelines are discussed.
The efficacy of the proposed technique is demonstrated in simulation for a two-tank system that is linearized along a time-varying reference trajectory.
 \end{abstract}

\end{frontmatter}

\begingroup
\def\QED{$\blacksquare$}
\let\oldepsilon\epsilon
\let\oldphi\phi
\renewcommand{\epsilon}{\varepsilon}
\renewcommand{\phi}{\varphi}
\newcommand{\ee}{\mathrm{e}}

\newcommand{\abs}[1]{\left\lvert #1 \right\rvert}
\newcommand{\norm}[2][]{\left\lVert #2 \right\rVert_{#1}}

\newcommand{\diffd}{\mathrm{d}}
\newcommand{\dt}[1]{\deriv{#1}{t}}
\newcommand{\deriv}[2]{\frac{\diffd {#1}}{\diffd  {#2}}}
\newcommand{\derivk}[3]{\frac{\diffd^{#3} {#1}}{\diffd  {#2}^{#3}}}
\newcommand{\pderiv}[2]{\frac{\partial{#1}}{\partial{#2}}}
\newcommand{\pderivs}[3]{\frac{\partial^2{#1}}{\partial{#2}\partial{#3}}}
\newcommand{\pderivk}[3]{\frac{\partial^{#3} {#1}}{\partial{#2}^{#3}}}

\newcommand{\integ}[4]{\int_{#2}^{#3}{{#4} \, \diffd {#1}}}
\newcommand{\integeq}[4]{\int_{{#1}={#2}}^{#3}{{#4} \, \diffd {#1}}}

\newcommand{\sbrm}[1]{\sb{\mathrm{#1}}}

\newcommand{\TT}{^{\mathrm{T}}}
\newcommand{\HH}{^{\mathrm{H}}}

\newcommand{\Matlab}{\textsc{Matlab}\xspace}
\newcommand{\Simulink}{\textsc{Simulink}\xspace}
\newcommand{\Labview}{\textsc{LabView}\xspace}

\def\A{\mathbf{A}}
\def\B{\mathbf{B}}
\def\C{\mathbf{C}}
\def\D{\mathbf{D}}
\def\E{\mathbf{E}}
\def\F{\mathbf{F}}
\def\G{\mathbf{G}}
\def\H{\mathbf{H}}
\def\I{\mathbf{I}}
\def\J{\mathbf{J}}
\def\K{\mathbf{K}}
\def\L{\mathbf{L}}
\def\M{\mathbf{M}}
\def\N{\mathbf{N}}
\def\O{\mathbf{O}}
\def\P{\mathbf{P}}
\def\Q{\mathbf{Q}}
\def\R{\mathbf{R}}
\def\S{\mathbf{S}}
\def\T{\mathbf{T}}
\def\U{\mathbf{U}}
\def\V{\mathbf{V}}
\def\W{\mathbf{W}}
\def\X{\mathbf{X}}
\def\Y{\mathbf{Y}}
\def\Z{\mathbf{Z}}

\def\a{\mathbf{a}}
\def\b{\mathbf{b}}
\def\c{\mathbf{c}}
\def\d{\mathbf{d}}
\def\e{\mathbf{e}}
\def\f{\mathbf{f}}
\def\g{\mathbf{g}}
\def\h{\mathbf{h}}
\def\i{\mathbf{i}}
\def\j{\mathbf{j}}
\def\k{\mathbf{k}}
\def\l{\mathbf{l}}
\def\m{\mathbf{m}}
\def\n{\mathbf{n}}
\def\o{\mathbf{o}}
\def\p{\mathbf{p}}
\def\q{\mathbf{q}}
\def\r{\mathbf{r}}
\def\s{\mathbf{s}}
\def\t{\mathbf{t}}
\def\u{\mathbf{u}}
\def\v{\mathbf{v}}
\def\w{\mathbf{w}}
\def\x{\mathbf{x}}
\def\y{\mathbf{y}}
\def\z{\mathbf{z}}

\newcommand{\valpha}{\bm{\alpha}}
\newcommand{\vbeta}{\bm{\beta}}
\newcommand{\vgamma}{\bm{\gamma}}
\newcommand{\vdelta}{\bm{\delta}}
\newcommand{\vepsilon}{\bm{\epsilon}}
\newcommand{\vzeta}{\bm{\zeta}}
\newcommand{\veta}{\bm{\eta}}
\newcommand{\vtheta}{\bm{\theta}}
\newcommand{\viota}{\bm{\iota}}
\newcommand{\vkappa}{\bm{\kappa}}
\newcommand{\vlambda}{\bm{\lambda}}
\newcommand{\vmu}{\bm{\mu}}
\newcommand{\vnu}{\bm{\nu}}
\newcommand{\vxi}{\bm{\xi}}
\newcommand{\vpi}{\bm{\pi}}
\newcommand{\vrho}{\bm{\rho}}
\newcommand{\vsigma}{\bm{\sigma}}
\newcommand{\vtau}{\bm{\tau}}
\newcommand{\vupsilon}{\bm{\upsilon}}
\newcommand{\vphi}{\bm{\phi}}
\newcommand{\vchi}{\bm{\chi}}
\newcommand{\vpsi}{\bm{\psi}}
\newcommand{\vomega}{\bm{\omega}}
\newcommand{\vAlpha}{\bm{\Alpha}}
\newcommand{\vBeta}{\bm{\Beta}}
\newcommand{\vGamma}{\bm{\Gamma}}
\newcommand{\vDelta}{\bm{\Delta}}
\newcommand{\vEpsilon}{\bm{\Epsilon}}
\newcommand{\vZeta}{\bm{\Zeta}}
\newcommand{\vEta}{\bm{\Eta}}
\newcommand{\vTheta}{\bm{\Theta}}
\newcommand{\vIota}{\bm{\Iota}}
\newcommand{\vKappa}{\bm{\Kappa}}
\newcommand{\vLambda}{\bm{\Lambda}}
\newcommand{\vMu}{\bm{\Mu}}
\newcommand{\vNu}{\bm{\Nu}}
\newcommand{\vXi}{\bm{\Xi}}
\newcommand{\vPi}{\bm{\Pi}}
\newcommand{\vRho}{\bm{\Rho}}
\newcommand{\vSigma}{\bm{\Sigma}}
\newcommand{\vTau}{\bm{\Tau}}
\newcommand{\vUpsilon}{\bm{\Upsilon}}
\newcommand{\vPhi}{\bm{\Phi}}
\newcommand{\vChi}{\bm{\Chi}}
\newcommand{\vPsi}{\bm{\Psi}}
\newcommand{\vOmega}{\bm{\Omega}}

\newcommand{\RR}{\mathbb{R}}
\newcommand{\CC}{\mathbb{C}}

\newcommand{\lambdamin}{\lambda\sbrm{min}}
\newcommand{\lambdamax}{\lambda\sbrm{max}}
\newcommand{\Ki}{\K\sbrm{I}}
\newcommand{\ki}{k\sbrm{I}}
\def\us{\u^{*}}

\section{Introduction}

The design of controllers containing open-loop integrators is one of the main tools for reducing the impact of slowly varying disturbances.
When considering set-point regulation of time-invariant plants, for example, such controllers can completely reject the influence of small parameter deviations and constant disturbances in steady state.
Techniques for designing such controllers have therefore seen extensive research, and some of the basic techniques are nowadays part of a typical curriculum on control systems, see e.g. \cite{franklin1986feedback}.

\changed{As an alternative to integral control, disturbance observer based control has seen extensive research, see, for example, the recent reviews by \cite{bakhshande2015proportional,chen2016disturbance,sariyildiz2020disturbance} and references therein.
In this framework, the disturbance is estimated by an observer and then cancelled by the control law.
The connection to integral control is the well-known internal model principle: with integral controllers, the integral part implicitly reconstructs the disturbance, while disturbance observers contain an explicit disturbance model.

In the time-varying case, rejection of disturbances by means of integral control or disturbance observers is more challenging.
For example, even constant disturbances can only be reconstructed if their time-varying influence on the plant is accurately known.
Moreover, their cancellation typically requires the disturbances to be matched, i.e., to act in the same channel as the control inputs.
In practice, time-varying dynamics often occur when linearizing nonlinear plants along time-varying references, cf. e.g., \cite{shao2014anovel}.
The possible performance improvement in this case, also in presence of more general (e.g., slowly varying) disturbances makes the use of integral-type controllers desirable in practice.

The present paper proposes an extension of a given time-varying state-feedback control law by an integral part while preserving its performance in the nominal case.
A similar goal may also be achieved by extending an existing observer by a disturbance estimate.
Such an extension, however, may require redesign or retuning of the entire observer.
Moreover, when the modeled disturbance is not matched, and hence cannot be cancelled straightforwardly, obtaining the final control law typically requires further effort.
The proposed approach, in contrast, requires little additional design effort, is easy to tune by means of an integral gain satisfying only a lower (but no upper) bound, and---being a standard integral control---is straightforward to add to almost any control architecture without significant modifications.

While the design of integral controllers is very well studied in general, there are only few explicit works on the time-varying case.
Most recent works consider it in the context of trajectory linearization control, see \cite{shao2014anovel,qiu2019robust}, and apply established techniques, such as feedback linearization, cf. \cite{palanki1997controller}, or time-varying LQR designs using an extended state space as in \cite{athans2013optimal,kalman1960contributions}, see also e.g. \cite{attia2020decentralized}.
These techniques do not allow to easily preserve the performance of a nominal controller, however.
The corresponding design of disturbance observers (sometimes called PI observers in case of constant disturbances) also is considered in the time-invariant case mostly; some of the few works on the time-varying case are found in \cite{kaczorek1979proportional,kaczorek1980reduced,shafai1985design}.
That design is based on a transformation to a normal form, however, which requires differentiability of the system coefficients. 
More recent works, such as \cite{ichalal2015onunknown,do2020robust,chen2020sliding}, handle time-varying systems in a linear parameter-varying (LPV) framework.
This typically leads to conservative designs, however, as the obtained controllers typically are valid for arbitrary parameter variations within a convex set.}

The approach presented in this paper is based on an idea recently proposed for the time-invariant, single-input single-output case in \cite{seeber2020performance}.
Here, this idea is extended to the time-varying, multivariable case.
Moreover, issues relevant in practice are addressed in the form of tuning insight and a modification to prevent integrator windup.

The paper is structured as follows:
After some preliminaries in Section~\ref{sec:prelim}, Section~\ref{sec:problem} introduces the problem statement and discusses the class of disturbances to be rejected.
The performance preserving integral state-feedback controller is then proposed in Section~\ref{sec:main}, and conditions for disturbance rejection and asymptotic stability are given.
The actual stability analysis and the corresponding proofs are contained in Section~\ref{sec:stability}.
Section~\ref{sec:implementation} discusses various issues that may be useful in a practical context: a modification to mitigate windup in the presence of limited control inputs, and an insight into the tuning of controller parameters.
Two special cases, the case of output-feedback integral action and the time-invariant case, are discussed in Section~\ref{sec:specialcases}.
Section~\ref{sec:simulation}, finally, applies the proposed approach \changed{in simulation to a two-tank system linearized along a time-varying reference trajectory.}
Section~\ref{sec:conclusion} draws conclusions.

\section{Preliminaries}
\label{sec:prelim}

This section discusses some notational conventions and stability notions that are used throughout the paper.
Matrices and vectors are denoted by boldface capital and boldface lowercase letters, respectively.
The largest and smallest eigenvalue of a symmetric matrix $\M = \M\TT$ are denoted by $\lambdamax(\M)$ and $\lambdamin(\M)$, respectively, the identity matrix is denoted by~$\I$, and $\norm{\v}$ or $\norm{\M}$ mean the (induced) 2-norm of a vector $\v$ or a matrix $\M$.

In dynamical systems, differentiation of a vector $\x$ with respect to time $t$ is expressed as $\dot \x$.
When writing such systems, time dependence of state (usually $\x$) and input (usually $\u$ or $\w$) is suppressed and only time dependence of the system's parameters is stated explicitly.
Furthermore, all time-varying system parameters are assumed to be uniformly bounded with respect to time.

Some stability notions for linear time-varying systems are discussed next.
For systems without inputs, the notions of asymptotic stability and uniform exponential stability, see, e.g. \cite{anderson2013stabilizability}, will be relevant.
\begin{defn}
    \label{def:stability}
    The linear system \mbox{$\dot \x = \A(t) \x$} is called
    \begin{enumerate}
        \item
            \emph{asymptotically stable (AS)}, if its origin is Lyapunov stable and every solution $\x(t)$ satisfies $\lim_{t \to \infty} \x(t) = \bm{0}$;
        \item
            \emph{uniformly exponentially stable (UES)}, if there exist positive constants $\mu$ and $M$ such that
            \begin{equation}\label{eq:ues_bound}
                \norm{\x(t)} \le M \ee^{-\mu (t - t_0)} \norm{\x(t_0)}
            \end{equation}
            holds for every solution $\x(t)$ and every $t_0$.
    \end{enumerate}
\end{defn}

Uniform exponential stability provides some robustness in the presence of bounded, vanishing disturbances:
\begin{lem}[{\cite[Theorem 59.1]{hahn1967stability}}]
    \label{th:hahn}
    Let $\dot \x = \A(t) \x$ be uniformly exponentially stable and consider the perturbed system $\dot \x = \A(t) \x + \w(t)$.
    If the disturbance $\w(t)$ is uniformly bounded and satisfies $\lim_{t \to \infty} \w(t) = \bm{0}$, then $\lim_{t \to \infty} \x(t) = \bm{0}$ holds for the perturbed system.
\end{lem}

If the initial state is set to zero, the boundedness of the system's state in the presence of a bounded input is guaranteed by uniform bounded-input, bounded-state stability as introduced in the following\footnote{This is a special case of uniform bounded-input, bounded-output stability as introduced in~\cite[Def. 12.1]{rugh1995linear}.}.

\begin{defn}
    The linear system \mbox{$\dot \x = \A(t) \x + \B(t) \w$} with input $\w$ is called \emph{uniformly bounded-input bounded-state stable} with respect to $\w$, if there exists a finite positive constant $\gamma$ such that for any $t_0$ and any input signal $\w(t)$, the corresponding zero-state response satisfies
	\begin{equation}
		\sup_{t\geq t_0} \norm{\x(t)} \le \gamma \sup_{t\geq t_0} \norm{\w(\tau)}.
	\end{equation}
\end{defn}
It is well known that for a uniformly bounded matrix $\B(t)$, uniform exponential stability of the autonomous system guarantees uniform bounded-input bounded-state stability, see, e.g.~\cite[Lemma 12.4]{rugh1995linear}.

\section{Problem Statement}
\label{sec:problem}
\newcommand{\zref}{\z\sbrm{ref}}
\newcommand{\uref}{\u\sbrm{ref}}
\newcommand{\yref}{\y\sbrm{ref}}

Consider a linear time-varying plant
\begin{subequations}
    \label{eq:plant}
    \begin{align}
        \label{eq:plant:xdot}
        \dot \x &= \A(t) \x + \B(t) \u + \F(t) \w \\
        \y &= \C(t) \x
    \end{align}
\end{subequations}
with a fully measurable state vector $\x \in \RR^n$, a control input $\u \in \RR^{l}$, an output $\y \in \RR^m$, and an external disturbance $\w \in \RR^{p}$.
The matrices $\A(t)$, $\B(t), \F(t), \C(t)$ are piecewise continuous and uniformly bounded with respect to~$t$.

The goal is to design a control law that asymptotically stabilizes the perturbed plant for a certain class of disturbances in the sense of the following definition:
\begin{defn}
    A control law is said to \emph{asymptotically stabilize} the perturbed plant \eqref{eq:plant} with respect to a certain class of perturbations, if the closed loop is asymptotically stable for $\w = \bm{0}$ and $\lim_{t \to \infty} \x(t) = \bm{0}$ holds for trajectories with all perturbations from the class.
\end{defn}

The design proposed in this paper is based on extending a given nominal (static) state-feedback controller by an integral action such that 
\begin{enumerate}
    \item
        the perturbed plant is asymptotically stabilized for a certain class of perturbations (specified below),
    \item 
        the plant is rendered uniformly bounded-input bounded-state stable with respect to arbitrary perturbations, and
    \item
        the nominal performance is preserved in the sense that, in the unperturbed case, behavior is identical to that obtained with the nominal controller.
\end{enumerate}

\newcommand{\qref}{\q\sbrm{ref}}
\changed{\begin{rem}
        Note that reference tracking for a linear plant $\dot \z = \A(t) \z + \B(t) \q + \F(t) \w$ can also be reduced to a stabilization problem as considered above\footnote{The use of the proposed approach for reference tracking is also demonstrated in the simulation example in Section~\ref{sec:simulation}.}.
        To see that, let $\zref(t)$ and $\qref(t)$ be a reference trajectory and input satisfying $\dot \z\sbrm{ref}(t) = \A(t) \zref(t) + \B(t) \qref(t)$.
Then, system \eqref{eq:plant:xdot} is obtained by choosing $\x = \z - \zref(t)$ and $\q = \u + \qref(t)$.
            Alternatively, if the reference input $\qref$ is unknown, one can also use $\q = \u$ and augment the disturbance vector $\w$ by $\qref$.
\end{rem}}

In the following, the assumptions regarding the nominal controller and the considered class of disturbances are discussed.

\subsection{Nominal State-Feedback Controller}

As a starting point for the considered controller design, a (static)  \changed{time-varying} state-feedback controller of the form
\begin{equation}
    \label{eq:ctrlnom}
    \u = -\K(t) \x
\end{equation}
with a uniformly bounded $\K(t) \in \RR^{l \times n}$ is assumed to be given.
It is assumed that the given state feedback renders the \emph{unperturbed} plant, i.e., \eqref{eq:plant:xdot} with $\w = \bm{0}$, uniformly exponentially stable.
This leads to the following formal assumption regarding the nominal closed loop.
\begin{assum}
    \label{ass:nominal}
    The state-feedback gain $\K(t)$ is such that the unperturbed, nominal closed-loop system 
    \begin{equation}
        \label{eq:nominal}
        \dot \x = \bigl[ \A(t) - \B(t) \K(t) \bigr] \x
    \end{equation}
    is uniformly exponentially stable.
\end{assum}
\begin{rem}
    Such a nominal controller may, for example, be designed by solving a Riccati differential equation, see e.g. \cite{kalman1960contributions}, \changed{or by means of spectrum assignment techniques, see e.g. \cite{zhu1997PD,babiarz2021necessary}}.
\end{rem}

\subsection{Disturbance Class for Asymptotic Stabilization}

In the time-invariant case, it is well-known that controllers with integral action can only compensate for (asymptotically) constant disturbances.
In the time-varying case, not the disturbance itself, but its action with respect to the control input needs to be constant.
This leads to the following specification of disturbances, for which asymptotic stabilization of the plant can reasonably be expected.
\begin{defn}
    \label{def:constant}
    A bounded disturbance $\w(t)$ is called \emph{asymptotically constant with respect to the control input}, if there exists a constant vector $\overline \w_0$ such that
    \begin{equation}
        \lim_{t \to \infty} \bigl[ \F(t) \w(t) - \B(t) \overline \w_0 \bigr] = \bm{0}
    \end{equation}
    holds.
\end{defn}

An important special case occurs when the disturbance is matched or asymptotically matched in the sense of the following definitions:
\begin{defn}
    The disturbance in \eqref{eq:plant} is called \emph{matched} or \emph{asymptotically matched}, respectively, if there exists a piecewise continuous and uniformly bounded matching matrix $\D(t)$ such that $\F(t) - \B(t) \D(t) = \bm{0}$ holds either for all $t$ or as $t \to \infty$.
\end{defn}
Clearly, a matched disturbance is also asymptotically matched.
In these cases, the considered class of disturbances may be specified in a simpler way using the following proposition:
\begin{prop}
    Consider system \eqref{eq:plant} and suppose that the disturbance is  bounded and asymptotically matched with matching matrix $\D(t)$.
    Then, $\w(t)$ is asymptotically constant with respect to the control input if $\lim_{t \to \infty} \D(t) \w(t)$ exists.
\end{prop}
\begin{proof}
    Set $\overline \w_0 = \lim_{t \to \infty} \D(t) \w(t)$ and note that
    \begin{align}
    \F(t) \w(t) - \B(t) \overline \w_0 &= \bigl( \F(t) - \B(t) \D(t) \bigr) \w(t)  \nonumber \\
    &\quad + \B(t) \bigl( \D(t) \w(t) - \overline \w_0 \bigr).
    \end{align}
    Since $\F(t) - \B(t) \D(t)$ and $\D(t) \w(t) - \overline \w_0$ tend to zero while $\w(t)$ and $\B(t)$ are bounded, one concludes
\begin{align}
    \lim_{t \to \infty} \F(t) \w(t) - \B(t) \overline \w_0 = \bm{0},
\end{align}
which completes the proof.
\end{proof}

\section{Performance Preserving Integral Control}
\label{sec:main}

The main results of this contribution---a performance preserving integral controller along with a stability condition---are presented in the following.

\subsection{Proposed Control Law}
\label{sec:main:state}

The proposed integral state-feedback control law is
\begin{subequations}
    \label{eq:ctrl}
    \begin{align}
        \u &= -\bigl[ \K(t) + \Ki \H(t) \bigr] \x + \Ki \v \\
        \label{eq:ctrl:v}
        \dot \v &= \G(t) \x,
    \end{align}
    where $\G(t)$ is calculated as
\begin{equation}
    \label{eq:ctrl:G}
    \G(t) = \dot \H(t) + \H(t) \bigl[ \A(t) - \B(t) \K(t) \bigr].
\end{equation}
\end{subequations}
Therein, a constant matrix $\Ki \in \RR^{l \times l}$ and the uniformly bounded feedback matrix $\H(t) \in \RR^{l \times n}$ with uniformly bounded time derivative $\dot \H(t)$ appear as parameters.
A structural representation of the proposed control law is shown in Fig.~\ref{fig:structure}.
Note that the output $\y$ is not used in \eqref{eq:ctrl}; the use of $\y$ for designing an output-feedback integral action \changed{and the problem of obtaining a controller in pure output-feedback are} discussed in Section~\ref{sec:output}.
\begin{figure}[tbp]
    \centering
    \includegraphics{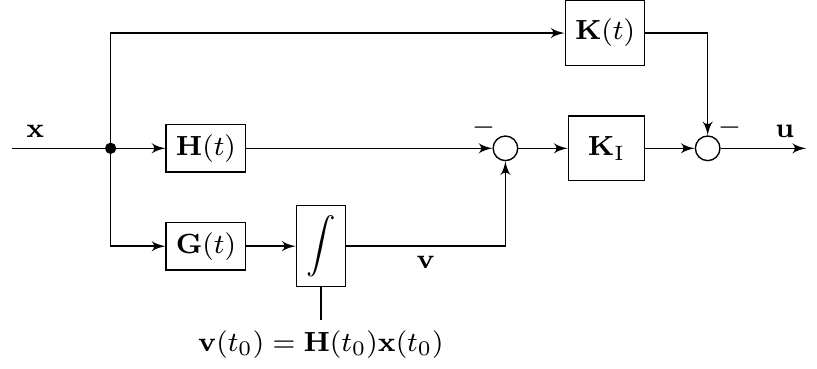}
    \caption{Block diagram of the proposed performance preserving integral control law \eqref{eq:ctrl} with tuning parameters $\H(t)$ and $\Ki$, nominal state-feedback matrix $\K(t)$, and abbreviation $\G(t)$ given in \eqref{eq:ctrl:G}.}
    \label{fig:structure}
\end{figure}

If the integrator's initial condition is chosen as
\begin{equation}
    \label{eq:ctrl:init}
    \v(t_0) = \H(t_0)\x(t_0),
\end{equation}
then the control law \eqref{eq:ctrl} preserves the performance of the nominal controller \eqref{eq:ctrlnom} in the unperturbed case; this is shown in the following proposition:
\begin{prop}
    \label{prop:performance}
    Consider the application of the control law \eqref{eq:ctrl} to the unperturbed plant, i.e., to \eqref{eq:plant:xdot} with \mbox{$\w(t) = \bm{0}$}.
    If the inital conditions of $\v$ and $\x$ satisfy \eqref{eq:ctrl:init}, then $\u(t) = -\K(t) \x(t)$ holds for all $t \ge t_0$.
\end{prop}
The proof of Proposition~\ref{prop:performance} is given in Section~\ref{sec:stability:closed}.
\begin{rem}
    \label{rem:performance}
    To see this result also intuitively, note that $\G(t) \x$ is the time derivative of $\H(t) \x$ along the plant's trajectories for $\u = -\K(t) \x$ and $\w = \bm{0}$.
    Therefore, with proper initialization, $\v(t) = \H(t) \x$ holds for all $t \ge t_0$ in this case, and only the nominal control remains.
\end{rem}

\subsection{Stability Condition}
\label{sec:main:stability}

In order to formulate a stability condition for the closed loop, the abbreviation
\begin{equation}
    \label{eq:def:Q}
    \Q(t) = \H(t) \B(t)
\end{equation}
is introduced.
Asymptotic stability of the overall closed loop may then be guaranteed using the following theorem, which is proven in Section~\ref{sec:stability:matched}.
\begin{thm}
    \label{th:stability}
    Consider the control law \eqref{eq:ctrl} with a symmetric parameter matrix $\Ki \in \RR^{l \times l}$ and a uniformly bounded feedback matrix $\H(t) \in \RR^{l\times n}$ with uniformly bounded time derivative $\dot \H(t)$.
    Let $\Q(t)$ be defined as in \eqref{eq:def:Q} and suppose that the nominal state feedback $\K(t)$ fulfills Assumption~\ref{ass:nominal}. 
    If $\Q(t) + \Q(t)\TT$ is positive semidefinite for every $t$ and there exist positive constants $\alpha$, $\beta$, and \changed{a non-negative constant} $T$ such that the inequalities
    \begin{subequations}
        \label{eq:cond}
        \begin{align}
\label{eq:cond:Ki}
            \lambdamin(\Ki) &\ge \alpha, \\
            \label{eq:cond:H}
            \integ{\sigma}{t_0}{t_0 + \tau}{\lambdamin\bigl[ \Q(\sigma) + \Q(\sigma)\TT \bigr]} &\ge 2 \beta \tau
        \end{align}
    \end{subequations}
    hold for all $t_0$ and all $\tau \ge T$, then the control law \eqref{eq:ctrl} asymptotically stabilizes the plant \eqref{eq:plant} for all perturbations that are asymptotically constant with respect to the control input in the sense of Definition~\ref{def:constant}.
\end{thm}
    The proof is given in Section~\ref{sec:stability:matched}.
\begin{rem}
    Note that the conditions on $\Ki$ and $\H(t)$ are decoupled.
    Once a time-varying feedback matrix $\H(t)$ satisfying the conditions is fixed, the proposed control law asymptotically stabilizes the perturbed plant for \emph{every} positive definite parameter matrix $\Ki$.
\end{rem}
\changed{\begin{rem}
    It is remarkable that the integral gain $\Ki$ can in fact be chosen arbitrarily large, as long as it is positive definite.
    This is a consequence of the fact that it affects $\v - \H(t) \x$ rather than just the integral state $\v$; for positive definite $\Q(t)$ as in \eqref{eq:def:Q}, the former is an output with relative degree one with respect to $\u$, which intuitively explains the lack of an upper bound for $\Ki$.
\end{rem}}

The following corollary presents a useful candidate for the choice of the time-varying feedback matrix $\H(t)$ with a simplified (though slightly more conservative) stability condition.
Further considerations and insights into the tuning of $\H(t)$ and $\Ki$ are presented in Section~\ref{sec:tuning}.
\begin{cor}
    \label{cor:stability}
    Suppose that $\dot \B(t)$ is uniformly bounded.
    If the symmetric parameter matrix $\Ki \in \RR^{l \times l}$ and $\B(t)$ satisfy, \changed{for positive constants $\alpha$ and $\beta$}, the conditions
    \begin{subequations}
        \label{eq:cond:simple}
        \begin{align}
            \lambdamin(\Ki) &\ge \alpha, \\
            \lambdamin(\B(t)\TT \B(t)) &\ge \beta
        \end{align}
    \end{subequations}
    for all $t$, then the control law \eqref{eq:ctrl} with $\H(t) = \B(t)\TT$ asymptotically stabilizes the plant \eqref{eq:plant} for all perturbations that are asymptotically constant with respect to the control input in the sense of Definition~\ref{def:constant}.
\end{cor}
\changed{\begin{rem}
        \label{rem:corstability}
        Note that, for $\H(t) = \B(t)\TT$, the inequalities \eqref{eq:cond:simple} imply conditions \eqref{eq:cond} of Theorem~\ref{th:stability} with $T = 0$ and the same constants $\alpha$, $\beta$.
\end{rem}}

\subsection{Bounded-Input Bounded-State Stability}

The previous considerations only guarantee stability for perturbations that are asymptotically constant with respect to the control input in the sense of Definition~\ref{def:constant}.
For arbitrary perturbations, bounded-input bounded-state stability of the closed loop may be shown under the same conditions:

\begin{thm}\label{th:bibs}
    Consider the closed loop obtained by applying control law \eqref{eq:ctrl} to the plant \eqref{eq:plant}.
    Suppose that the conditions of Theorem~\ref{th:stability} are fulfilled; in particular, let system \eqref{eq:nominal} be uniformly exponentially stable with positive constants $\mu$ and $M$  \changed{as in Definition~\ref{def:stability}} and let $B$, $F$, $H$ be uniform upper bounds for $\norm{\B(t)}$, $\norm{\F(t)}$, $\norm{\H(t)}$, respectively.
Then, the \emph{closed loop} is uniformly bounded-input bounded-state stable with respect to the perturbation input $\w$.
    In particular, the \emph{controlled plant} is also rendered uniformly bounded-input bounded-state stable with the gain
    \begin{equation}\label{eq:gamma}
        \gamma = \frac{B M}{\mu} \ee^{\alpha \beta T} \sqrt{\frac{\lambdamax(\Ki)}{\lambdamin(\Ki)}} \frac{\lambdamax(\Ki)}{\alpha} \frac{H F}{\beta}+\frac{F M}{\mu}.
    \end{equation}
\end{thm}
The proof is given in Section~\ref{sec:stability:bibo}.

\section{Stability Analysis}
\label{sec:stability}

In this section, the proposed approach for designing an integral controller is discussed in more detail.
First, the closed-loop system is derived and the performance preserving property in Proposition~\ref{prop:performance} is shown.
Stability is then analyzed, and the stability condition in Theorem~\ref{th:stability} and the bounded-input bounded-state gain of Theorem~\ref{th:bibs} are proven.

\subsection{Closed Loop System}
\label{sec:stability:closed}

In order to give a state-space representation of the closed loop, the state variable
\begin{equation}
    \label{eq:z}
    \z = - \Ki \H(t) \x + \Ki \v + \w_0
\end{equation}
with a constant vector $\w_0$ (to be defined later), and the abbreviation
\begin{equation}\label{eq:wtilde}
    \tilde \w(t) = \F(t) \w(t) - \B(t) \w_0
\end{equation}
are introduced.

According to \eqref{eq:plant} and \eqref{eq:ctrl}, the closed loop is then governed by\footnote{Note that $\G(t)$ is the time derivative of $\H(t)$ along the trajectories of the nominal closed loop $\dot \x = \bigl[ \A(t) - \B(t) \K(t) \bigr] \x$.}
\begin{subequations}
    \label{eq:closed}
    \begin{align}
        \label{eq:closed:x}
        \dot \x &= \bigl[ \A(t) - \B(t) \K(t) \bigr] \x + \B(t) \z + \tilde \w \\
        \label{eq:closed:z}
        \dot \z &= - \Ki \H(t) \bigl[ \B(t) \z + \tilde \w \bigr].
    \end{align}
\end{subequations}

In the unperturbed case, i.e., for $\w(t) = \w_0 = \bm{0}$ and hence $\tilde \w(t) = \bm{0}$, the second of these equations reduces to
\begin{equation}
    \label{eq:closed:unperturbed:z}
    \dot \z = - \Ki \H(t) \B(t) \z,
\end{equation}
and the control input is given by
\begin{equation}
    \label{eq:closed:unperturbed:u}
    \u = - \K(t) \x + \z.
\end{equation}
Using these considerations, Proposition~\ref{prop:performance} may be proven.

\textit{Proof of Proposition~\ref{prop:performance}:}
Initial condition \eqref{eq:ctrl:init} and \mbox{$\w(t_0) = \bm{0}$} imply $\z(t_0) = \bm{0}$.
Therefore, $\z(t) = \bm{0}$ is the unique solution of \eqref{eq:closed:unperturbed:z}, and \eqref{eq:closed:unperturbed:u} yields \mbox{$\u(t) = -\K(t) \x(t)$} for all $t \ge t_0$.
\hfill~\qed

\subsection{Asymptotic Stabilization}
\label{sec:stability:matched}

The stability of the closed-loop system is now studied for disturbances that are asymptotically constant with respect to the control input in the sense of Definition~\ref{def:constant}.
To that end, the vector $\w_0$ in the closed-loop description is set to $\overline{\w}_0$ from that definition, and hence $\tilde \w(t)$ vanishes asymptotically.

Using \eqref{eq:z} to obtain the representation \eqref{eq:closed} of this system preserves the closed-loop stability properties, if the associated state transformation
\begin{equation}
    \begin{bmatrix}
        \x \\
        \z
    \end{bmatrix} = \begin{bmatrix}
        \I & \bm{0} \\
        -\Ki \H(t) & \Ki
    \end{bmatrix} \begin{bmatrix}
        \x \\
        \v
    \end{bmatrix} + \begin{bmatrix}
        \bm{0} \\
        \w_0
    \end{bmatrix}
\end{equation}
is a Lyapunov transformation, i.e., if the transformation matrix
\begin{equation}
    \M(t) = \begin{bmatrix}
        \I & \bm{0} \\
        -\Ki \H(t) & \Ki
    \end{bmatrix}
\end{equation}
has a uniformly bounded time derivative $\dot \M(t)$ and inverse $\M(t)^{-1}$, see e.g.~\cite[Chapter III]{adrianova1995introduction}.
This is the case if $\dot \H$ is bounded and $\Ki$ is invertible.
As one can see from \eqref{eq:closed}, closed-loop stability is then determined by the stability of the subsystem \eqref{eq:closed:z} governing the state variable $\z$.
Since it is excited by a vanishing disturbance, uniform exponential stability of its autonomous part guarantees its asymptotic stability.
This yields the following intermediate result.
\begin{lem}
    \label{lem:zdot}
    Consider the closed-loop system \eqref{eq:wtilde},~\eqref{eq:closed} for disturbances $\w(t)$ that are asymptotically constant with respect to the control input in the sense of Definition~\ref{def:constant}.
Suppose that the nominal state feedback $\K(t)$ fulfills Assumption~\ref{ass:nominal} and that the system
    \begin{equation}
        \label{eq:zdot}
        \dot \z = -\Ki \H(t) \B(t) \z
    \end{equation}
    is uniformly exponentially stable.
    Then, the closed loop is asymptotically stable.
\end{lem}
\begin{proof}
	    Consider first system \eqref{eq:closed:z}, which governs the perturbed trajectories of $\z$.
        Since the corresponding unperturbed system \eqref{eq:zdot} is uniformly exponentially stable, Lemma~\ref{th:hahn} along with \mbox{$\lim_{t \to \infty} \tilde \w(t) = \bm{0}$} implies $\lim_{t \to \infty} \z(t) = \bm{0}$.

	Interpret now the remaining system dynamics \eqref{eq:closed:x} as a perturbed system with disturbances $\z(t)$ and $\tilde \w(t)$.
	Since both tend to zero asymptotically, and the unperturbed system is uniformly exponentially stable by virtue of Assumption~\ref{ass:nominal}, applying Lemma~\ref{th:hahn} again guarantees that also $\x(t)$ tends to zero asymptotically.
	This concludes the proof.
\end{proof}

Verifying the condition of Lemma~\ref{lem:zdot} is not easy in general.
A more useful stability condition may be obtained by considering the function $V(\z) = \z\TT \P \z$ with a positive definite matrix $\P$ as a quadratic candidate Lyapunov function for system \eqref{eq:zdot}.
Its time derivative $\dot V$ along the trajectories of \eqref{eq:zdot} is given by
\begin{equation}
    \dot V(t, \z) = -\z\TT \bigl[ \B(t)\TT \H(t)\TT \Ki\TT \P + \P \Ki \H(t) \B(t) \bigr] \z.
\end{equation}
One can see that by choosing $\P = \Ki^{-1}$ and selecting the controller parameter $\Ki$ as a symmetric, positive definite matrix, the resulting stability condition for $\H(t)$ can be decoupled from $\Ki$.
Using these considerations, Theorem~\ref{th:stability} can now be proven.

\textit{Proof of Theorem~\ref{th:stability}:}
In order to show uniform exponential stability of system \eqref{eq:zdot}, the Lyapunov function candidate $V(\z) = \z\TT \Ki^{-1} \z$ is considered.
    Its time derivative $\dot V$ along the trajectories of \eqref{eq:zdot} satisfies
\begin{align}
        \dot V(t,\z) &= -\z\TT \bigl[ \H(t) \B(t) + \B(t)\TT \H(t)\TT \bigr] \z \nonumber \\
        &\le -\lambdamin\bigl[ \H(t) \B(t) + \B(t)\TT \H(t)\TT \bigr] \z\TT \z.
    \end{align}
    Introducing $\gamma(t) = \lambdamin\bigl[ \H(t) \B(t) + \B(t)\TT \H(t)\TT \bigr] \ge 0$ as an abbreviation and using $\z\TT \Ki^{-1} \z \le \lambdamax(\Ki^{-1}) \z\TT \z$, one has
\begin{align}
        \dot V(t, \z) &\le -\frac{\gamma(t)}{{\lambdamax( \Ki^{-1} )}} V(\z) \le - \alpha \gamma(t) V(\z).
    \end{align}
    Integrating this inequality yields
    \begin{equation}
        \label{eq:V:ineq}
        V(\z(t)) \le \exp\Bigl(-\alpha \integ{\sigma}{t_0}{t}{\gamma(\sigma)}\Bigr) V(\z(t_0)).
    \end{equation}
According to \eqref{eq:cond:H}, $\gamma$ satisfies
    \begin{equation}
        \integ{\sigma}{t_0}{t}{\gamma(\sigma)} \ge \begin{cases}
            0 & t - t_0 < T \\
            2 \beta (t - t_0) & t - t_0 \ge T.
        \end{cases}
    \end{equation}
    Thus,
    \begin{equation}
        \integ{\sigma}{t_0}{t}{\gamma(\sigma)} \ge 2 \beta (t - t_0) - 2 \beta T
    \end{equation}
    holds in either of the two cases, and substitution into \eqref{eq:V:ineq} yields
    \begin{align}
        V(\z(t)) \le \ee^{2 \alpha \beta T} \ee^{- 2 \alpha \beta (t - t_0)} V(\z(t_0)).
    \end{align}
Since $\lambdamin(\Ki^{-1}) \norm{\z}^2 \le V(\z) \le \lambdamax(\Ki^{-1}) \norm{\z}^2$ holds, one obtains
    \begin{equation}
        \label{eq:z:exponential}
        \norm{\z(t)} \le \ee^{\alpha \beta T} \sqrt{\frac{\lambdamax(\Ki)}{\lambdamin(\Ki)}} \ee^{-\alpha \beta (t - t_0)} \norm{\z(t_0)}.
    \end{equation}
    This shows uniform exponential stability of system \eqref{eq:zdot}, and the proof is concluded by applying Lemma~\ref{lem:zdot}.
\hfill~\qed

\subsection{Bounded-Input Bounded-State Stability Gain}
\label{sec:stability:bibo}
In order to analyze the behavior for general disturbances, the closed loop is now studied for $\w_0 = \bm{0}$, i.e., with $\tilde \w = \F(t) \w$.
Theorem~\ref{th:bibs} may then be proven.

\textit{Proof of Theorem~\ref{th:bibs}:}
In the unperturbed case, i.e., $\tilde \w =\bm0$, the closed-loop system~\eqref{eq:closed} is uniformly exponentially stable.
This follows from \cite[Theorem 2]{zhou2016on} because~\eqref{eq:nominal} and \eqref{eq:closed:unperturbed:z} are uniformly exponentially stable and~\eqref{eq:closed} is in block triangular form.
According to~\cite[Lemma 12.4]{rugh1995linear}, this guarantees uniform bounded-input bounded-state stability.
The gain $\gamma$ for the closed-loop plant states as stated in~\eqref{eq:gamma} will be derived in the following.

Let $\bm\Phi(t,t_0)$ and $\bm\Phi\sbrm{z}(t,t_0)$ denote the state transition matrices of~\eqref{eq:nominal} and \eqref{eq:closed:unperturbed:z}, respectively.
Bounds for these transition matrices are given by
\begin{subequations}
\begin{align}
\norm{\bm\Phi(t,t_0)} &\leq Me^{-\mu (t-t_0)}	\quad\text{ and } \\
\norm{\bm\Phi\sbrm{z}(t,t_0)} &\leq M\sbrm{z}e^{-\mu\sbrm{z} (t-t_0)}
\end{align}
\end{subequations}
with $M,\,\mu >0$ as in the theorem and $\mu\sbrm{z} = \alpha\beta$ and $M\sbrm{z} =  \ee^{\alpha \beta T} \sqrt{\frac{\lambdamax(\Ki)}{\lambdamin(\Ki)}}$, obtained from~\eqref{eq:z:exponential}. 

For a general perturbation $\w(t)$ and with $\w_0 = \bm{0}$, the effect of the input $\w(t)$ on $\z(t)$ (with $\z(t_0)=\bm 0$) in system \eqref{eq:zdot} is given by
\begin{equation}
    \z(t)= -\integ{\tau}{t_0}{t}{\bm\Phi\sbrm{z}(t,\tau) \Ki \H(\tau)\F(\tau)\w(\tau)}.
\end{equation}
Using the upper bounds $B, H, F$ for $\B(t), \H(t), \F(t)$, one obtains the bound
\begin{align}\label{eq:bibo_z}
\|\z(t)\| &\leq \integ{\tau}{t_0}{t}{M\sbrm{z} e^{-\mu\sbrm{z}(t-\tau)} \norm{\Ki} H F} \sup_{\tau\in[t_0,t]}\norm{\w(\tau)} \nonumber \\
    &\leq \frac{M\sbrm{z}}{\mu\sbrm{z}} HF\lambdamax(\Ki)\sup_{\tau\in[t_0,t]}\norm{\w(\tau)}.
\end{align}
The zero state response of the plant state $\x(t)$ can be stated as
\begin{equation}\label{eq:bibo:zerostate}
	\x(t) = \integ{\tau}{t_0}{t}{\bm\Phi(t,\tau)\B(\tau)\z(\tau)}+ \integ{\tau}{t_0}{t}{\bm\Phi(t,\tau)\F(\tau)\w(\tau)}.
\end{equation}
Performing estimates analogous to~\eqref{eq:bibo_z} and using this bound in~\eqref{eq:bibo:zerostate} results in
\begin{align}
    \|\x(t)\|&\leq \frac{BM}{\mu}\frac{M\sbrm{z}}{\mu\sbrm{z}} HF\lambdamax(\Ki)\sup_{\tau\in[t_0,t]}\norm{\w(\tau)} \nonumber \\
	&\quad + \frac{FM}{\mu}\sup_{\tau\in[t_0,t]}\norm{\w(\tau)}.
\end{align}
Taking the supremum on the left hand side over \mbox{$t\geq t_0$} shows that the controlled plant is uniformly bounded-input bounded-state stable with gain~\eqref{eq:gamma}. \hfill~\qed

\section{Implementation Issues}
\label{sec:implementation}

This section discusses two practical aspects of the proposed control law: the mitigation of integrator windup and the choice of parameters.

\subsection{Mitigation of Windup}
\label{sec:windup}

In the presence of control input saturation, controllers with integral feedback are known to suffer from an effect called \emph{controller windup}, see e.g. \cite{hippe2006windup}.
While the control input is saturated, the internal state of the controller may wind up, causing large, undesired overshoots or even unbounded trajectories.
This section presents a way to mitigate this problem for the proposed control law.

Suppose that the control input, which is actually applied to the plant, is given by $\us$ rather than $\u$.
The signal $\us$ may be obtained from $\u$, for example, by component-wise saturation functions.
Here, the only assumption made about $\us$ is that $\us = \u$, when $\u$ satisfies the control input constraints.

In order to avoid windup, the control law \eqref{eq:ctrl} may be modified as
\begin{subequations}
    \label{eq:ctrl:aw}
    \begin{align}
        \u &= -\bigl[ \K(t) + \Ki \H(t) \bigr] \x + \Ki \v \\
        \label{eq:ctrl:aw:v}
        \dot \v &= \G(t) \x + \H(t) \B(t) ( \us - \u ).
    \end{align}
\end{subequations}
Fig.~\ref{fig:structure:aw} depicts a block diagram of this modified control law.
It is motivated by the desire to maintain the property pointed out in Remark~\ref{rem:performance} also in the case $\u \ne \us$: that the right-hand side of $\dot\v$ stays equal to the time derivative of $\H(t) \x$.

\begin{figure}[tbp]
    \centering
    \includegraphics{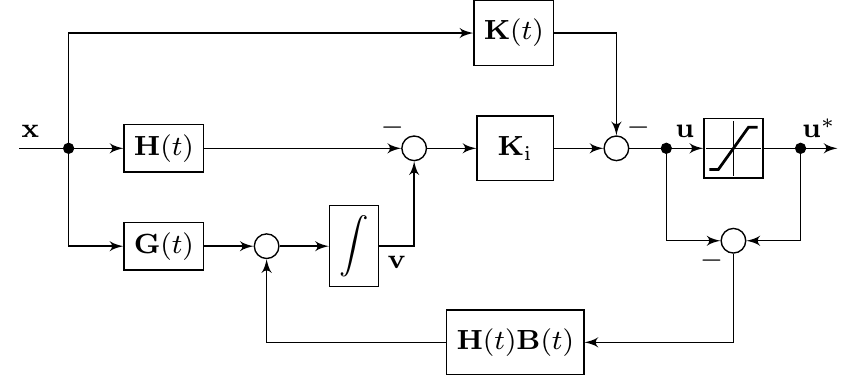}
    \caption{Block diagram of the proposed performance preserving integral control law with anti-windup \eqref{eq:ctrl:aw}, assuming a saturation nonlinearity between unconstrained control input $\u$ and constrained control input $\us$.}
    \label{fig:structure:aw}
\end{figure}

With this modification, the following asymptotic property of $\u(t)$ can be shown for the constrained closed loop.
\begin{prop}
    \label{prop:windup}
    Consider the plant \eqref{eq:plant} with the unconstrained control input $\u$ replaced by a constrained input $\us$.
    If the conditions of Theorem~\ref{th:stability} are fulfilled, then the unconstrained control input $\u(t)$ obtained with the control law \eqref{eq:ctrl:aw}, \eqref{eq:ctrl:G} satisfies
    \begin{equation}
        \lim_{t \to \infty} \u(t) + \K(t) \x(t) + \w(t) = \bm{0},
    \end{equation}
i.e., $\u(t)$ asymptotically tends to $-\K(t) \x(t) - \w(t)$.
\end{prop}
\begin{proof}
    One may verify that after the replacement of $\u$ by $\us$ in \eqref{eq:plant} and of \eqref{eq:ctrl:v} by \eqref{eq:ctrl:aw:v}, the variable $\z$ defined in \eqref{eq:z} still satisfies \eqref{eq:closed:z}.
    Therefore, the proof of Theorem~\ref{th:stability} is applicable without modification to show that $\z(t)$ tends to zero.
Noting that $\z(t) = \u(t) + \K(t) \x(t) + \w(t) - \tilde \w(t)$ and that $\tilde \w(t)$ tends to zero completes the proof.
\end{proof}
\begin{rem}
Note that although the nominal control signal with a disturbance compensation is recovered asymptotically, \emph{plant windup} or the \emph{directionality problem} may additionally occur in the presence of input saturation, see, e.g. \cite{hippe2006windup}.
    Therefore, no general formal statements about closed-loop stability in the presence of saturation nonlinearities can be made, but \emph{controller windup} of the integrator is prevented.
\end{rem}
\changed{\begin{rem}
        From this proof and from \eqref{eq:zdot}, one can see that nominal behavior is approached the faster, the larger $\Ki$ is.
    Hence, also in the perturbed case, nominal performance is recovered with increasing integrator gain.
\end{rem}}

\subsection{Tuning of Parameters}
\label{sec:tuning}

This section discusses some guidelines for the choices of $\Ki$ and $\H(t)$.
Regarding $\Ki$, one can see from the proofs in the previous section, in particular from \eqref{eq:z:exponential}, that the exponential convergence rate $\mu$ in the sense of Definition~\ref{def:stability} is given by
$
    \mu = \alpha \beta
$
with the positive constants $\alpha$ and $\beta$ from Theorem~\ref{th:stability}.
Along with that theorem's conditions, this suggests that a desired convergence rate $\mu^*$ can be ensured by selecting the positive definite controller parameter $\Ki$ such that
$
    \lambdamin(\Ki) \ge \frac{\mu^*}{\beta}.
$

As pointed out in Corollary~\ref{cor:stability}, one possible choice for $\H(t)$ is $\H(t) = \B(t)\TT$.
\changed{As pointed out in Remark~\ref{rem:corstability}}, $\beta$ has the same meaning as in Theorem~\ref{th:stability} \changed{in this case}, i.e., it may be used for tuning $\Ki$ as discussed before.
Under conditions of the corollary, $\H(t)$ may also be chosen as
\begin{equation}
    \label{eq:H:choice}
    \H(t) = \frac{\B(t)\TT}{\norm{\B(t)}^2} \quad \text{or} \quad
    \H(t) = \frac{\B(t)\TT}{\lambdamin(\B(t)\TT \B(t))}.
\end{equation}
The latter choice, in particular, achieves
\begin{equation}
    \integ{\sigma}{t_0}{t_0 + \tau}{\lambdamin\bigl[ \Q(\sigma) + \Q(\sigma)\TT \bigr]} = 2 \tau
\end{equation}
for all $\tau \ge 0$ and $t_0$, i.e., equality is obtained in condition \eqref{eq:cond:H} of Theorem~\ref{th:stability} with $\beta = 1$, which yields the convergence rate $\mu = \lambdamin(\Ki)$.

If $\norm{\B(t)}$ or $\lambdamin(\B(t)\TT \B(t))$ is not uniformly bounded from below by a positive constant, choosing $\H(t)$ is less straightforward.
In this case, the choice $\H(t) = \B(t)\TT$ or variants of \eqref{eq:H:choice} such as
\begin{equation}
    \H(t) = \frac{\B(t)\TT}{\max( \lambdamin(\B(t)\TT \B(t)), L )}
\end{equation}
with $L > 0$ may be explored, but in general $\H(t)$ has to be chosen in accordance with the conditions of Theorem~\ref{th:stability}, which have be checked on a case-to-case basis.

\section{Special Cases}
\label{sec:specialcases}

This section discusses two important special cases of the controller whose general form is given in \eqref{eq:ctrl}:
the design of a (time-varying) output integral feedback, and the design for a time-invariant plant.

\subsection{Output-Feedback Integral Action}
\label{sec:output}

In practice, it is sometimes desired that the integral controller should be designed using the integral of a given output $\y$.
The problem then becomes that of finding a time-varying gain $\M(t)$ and a state-feedback gain $\H(t)$ such that the control law \eqref{eq:ctrl} may be written as
\begin{subequations}
    \label{eq:ctrl:outputfeedback}
    \begin{align}
        \u &= -\bigl[ \K(t) + \Ki \H(t) \bigr] \x + \Ki \v \\
    \label{eq:ctrl:outputfeedback:vdot}
        \dot \v &= \M(t) \y,
    \end{align}
\end{subequations}
i.e., such that $\G(t) = \M(t) \C(t)$ holds in \eqref{eq:ctrl} for all $t$.
To fulfill \eqref{eq:ctrl:G}, $\H(t)$ then has to be a solution of the system
\begin{subequations}
    \label{eq:outputfeedback:H}
    \begin{align}
        \dot \H &= - \H \bigl[ \A(t) - \B(t) \K(t) \bigr] + \M \C(t), \\
        \Q &= \H \B(t).
    \end{align}
\end{subequations}
Therein, $\M \in \RR^{l \times m}$ acts as an input, $\H \in \RR^{l \times n}$ is the (matrix-valued) state, and the output $\Q \in \RR^{l \times l}$ is relevant for the stability condition in Theorem~\ref{th:stability}.

Finding a solution for this system can be interpreted as a control problem for the dual of the nominal closed loop.
To see this, the $i$-th rows of $\H$, $\M$, and $\Q$ are denoted by $\h_i \in \RR^{n}$, $\m_i \in \RR^{m}$, and $\q_i \in \RR^{l}$, respectively, i.e.,
\begin{equation}
    \H = \begin{bmatrix}
        \h_1\TT \\
        \vdots \\
        \h_l\TT
    \end{bmatrix}, \quad
    \M = \begin{bmatrix}
        \m_1\TT \\
        \vdots \\
        \m_l\TT
    \end{bmatrix}, \quad
    \Q = \begin{bmatrix}
        \q_1\TT \\
        \vdots \\
        \q_l\TT
    \end{bmatrix}.
\end{equation}
Substitution into \eqref{eq:outputfeedback:H} shows that the transposed rows are governed by the system
\begin{subequations}
    \label{eq:dual}
    \begin{align}
        \dot \h_i &= -\bigl[ \A(t) - \B(t) \K(t) \bigr]\TT \h_i + \C(t)\TT \m_i \\
        \q_i &= \B(t)\TT \h_i.
    \end{align}
\end{subequations}
This system is the dual of the nominal closed loop.
It is therefore anti-stable, i.e., uniformly exponentially stable in reverse time.
For bounded $\m_i(t)$, the existence of bounded solutions is guaranteed from the fact that the system has an exponential dichotomy, see, e.g., \cite[Ch. 3, Proposition 2]{coppel1978dichotomies}.
Depending on the structure of the system, such solutions with desired $\Q(t)$ may be found, for example, using flatness-based or input-output linearization techniques.

\changed{\begin{rem}
        Note that although the integral \eqref{eq:ctrl:outputfeedback:vdot} is computed only from the output, the overall control law \eqref{eq:ctrl:outputfeedback} still requires full-state feedback.
        To obtain a pure output feedback controller, an unknown input observer may be used to reconstruct the state $\x$ from the measured output $\y$ without knowledge of the disturbance $\w$, see e.g. \cite{ichalal2015onunknown,tranninger2021strong}.
\end{rem}}

\subsection{Time-Invariant Case}

Consider the time-invariant case, i.e., a time-invariant plant $\dot \x = \A \x + \B \u$ and nominal control law $\u = - \K \x$.
Then, the gain matrices $\H$ and $\G$ may also be chosen to be constant.
Considering, in particular, the output-feedback case, one may choose $\G = \M \C$ \changed{with constant matrix $\M$} and compute $\H$ according to \eqref{eq:ctrl:G} as
\begin{equation}
    \label{eq:H:timeinvariant}
    \H = \G ( \A - \B \K )^{-1} = \M \C ( \A - \B \K )^{-1}.
\end{equation}
In this case, the control law \eqref{eq:ctrl:outputfeedback} becomes
\begin{subequations}
    \label{eq:ctrl:timeinvariant}
    \begin{align}
        \u &= -\Bigl[ \K + \Ki \M \C ( \A - \B \K )^{-1} \Bigr] \x + \Ki \v \\
        \dot \v &= \M \y.
    \end{align}
\end{subequations}
A reasonable choice for the constant matrix $\M \in \RR^{l \times m}$ is given by the following proposition, which is a generalization of \cite[Proposition 1]{seeber2020performance} to the multivariable case.

\begin{prop}
    \label{prop:timeinvariant}
    Consider the closed loop formed by applying the control law \eqref{eq:ctrl:timeinvariant} to the time-invariant plant $\dot \x = \A \x + \B \u$, $\y = \C \x$ and suppose that the matrix \mbox{$\A - \B \K$} is Hurwitz.
    If $\C ( \A - \B \K )^{-1} \B$ is left-invertible and $\M$ is the corresponding left (pseudo-)inverse
    \begin{equation}
        \label{eq:M:timeinvariant}
        \M = \bigl[ \C ( \A - \B \K )^{-1} \B \bigr]^{+},
    \end{equation}
    then the closed-loop eigenvalues are given by the union of the eigenvalues of the matrices $\A - \B \K$ and $-\Ki$.
\end{prop}
\begin{proof}
    With the considered value of $\M$ and taking into account \eqref{eq:H:timeinvariant}, one has $\H \B = \M \C (\A - \B \K)^{-1} \B = \I$.
The unperturbed closed-loop system \eqref{eq:closed} hence is
\begin{align}
    \begin{bmatrix}
        \dot \x \\
        \dot \z
    \end{bmatrix}
&= \begin{bmatrix}
        \A - \B \K & \B  \\
        \bm{0} & -\Ki
    \end{bmatrix} \begin{bmatrix}
        \x \\ \z
    \end{bmatrix}.
\end{align}
The claimed statement is then obvious from the system's block triangular structure.
\end{proof}
\begin{rem}
    The performance preserving effect of the proposed controller, which is achieved by selecting the initial value according to Proposition~\ref{prop:performance} as
    \begin{equation}
        \v(t_0) = \M \C ( \A - \B \K )^{-1} \x(t_0),
    \end{equation}
    can here also be seen from the fact that the controller preserves the nominal closed-loop eigenvalues, while the additional eigenvalues may be tuned using $\Ki$.
\end{rem}
\begin{rem}
    Note that invertibility of $\C (\A - \B \K)^{-1} \B$ is a reasonable assumption, because it is equivalent to the absence of transmission zeros in the plant at zero, i.e., \changed{to a non-singular dc-gain.}
\end{rem}

\newcommand{\xeq}{\overline{x}}
\newcommand{\ueq}{\overline{u}}
\newcommand{\zrefi}[1]{z_{\mathrm{ref},#1}}
\newcommand{\qrefi}{q_{\mathrm{ref}}}
\newcommand{\zrefdoti}[1]{\dot{z}_{\mathrm{ref},#1}}
\section{Simulation Example}
\label{sec:simulation}

\changed{
The presented approach is demonstrated in a simulation using a two-tank system as considered, e.g., in \cite{pan2005experimental}.
The plant is goverened by the nonlinear model \mbox{$\dot z_1 = -c_1 \sqrt{z_1} + c_3 q + w$}, $\dot z_2 = c_1 \sqrt{z_1} - c_2\sqrt{z_2}$ with measured liquid levels $z_1, z_2$, pump voltage $q$, disturbance $w$, and positive parameters $c_1, c_2, c_3$.
The goal is for the lower tank level $z_2$ to track a given reference $r(t) = c_4 + c_5 \sin(2 \pi c_6 t)$ with positive constants $c_4 \ge c_5$ and $c_6$.
For control design, the system is linearized along the reference trajectory \mbox{$\zrefi{1}(t) = c_1^{-2} (c_2 \sqrt{r(t)} + \dot r(t))^2$}, $\zrefi{2}(t) = r(t)$, see, e.g., \cite[Chapter 5.2]{rudolph2021flatness} or \citep{shao2014anovel},
and the control input is chosen as $q = u + \qrefi(t)$ with $\qrefi(t) = c_3^{-1} \zrefdoti{1}(t) + c_1 c_3^{-1} \sqrt{\zrefi{1}(t)}$ to obtain the linear, time-varying dynamics in form \eqref{eq:plant:xdot}
\begin{equation}
    \label{eq:exmp:param}
    \dot \x = \begin{bmatrix}
        -\frac{c_1}{2 \sqrt{\zrefi{1}(t)}} & 0 \\
        \frac{c_1}{2 \sqrt{\zrefi{1}(t)}} & -\frac{c_2}{2 \sqrt{\zrefi{2}(t)}} \\
    \end{bmatrix} \x + \begin{bmatrix}
        c_3 \\
        0
    \end{bmatrix} u + \begin{bmatrix}
        1 \\
        0
    \end{bmatrix} w
\end{equation}
for the linearized tracking error $\x = \z - \zref(t)$.
The model parameters $c_1, c_2, c_3$ are taken from the experimental setup in \citep{pan2005experimental}; Table~\ref{tab:parameters} lists them along with parameters $c_4, c_5, c_6$ of the reference.

Due to the lower triangular structure and the chosen reference, system \eqref{eq:exmp:param} can be shown to be UES for \mbox{$u = w = 0$}, see \cite[Lemma 5 \& Theorem 2]{zhou2016on}.
Hence, the nominal state feedback $\K(t) = \bm{0}$ is used in the simulation example for simplicity.
Choosing, furthermore, $\H(t) = [\alpha \quad \alpha]$ with constant $\alpha > 0$, computing $\G(t)$ from \eqref{eq:ctrl:G} and \eqref{eq:exmp:param}, and taking into account the linearization, relation \eqref{eq:ctrl:aw} yields  the overall control law
\begin{align}
    \label{eq:sim:ctrl}
    q &= \qrefi(t) - \ki \alpha (z_1 - \zrefi{1}(t) + z_2 - \zrefi{2}(t)) + \ki v, \nonumber\\
\dot v &= - \frac{\alpha c_2 }{2 \sqrt{\zrefi{2}(t)}} (z_2 - \zrefi{2}(t))+ \alpha c_3 (q^* - q)
\end{align}
where $u^* = q^* - \qrefi(t)$ is substituted in \eqref{eq:ctrl:aw}, and \mbox{$q^* = \max(0, \min(Q, q))$} denotes the saturated input which is applied to the plant ($Q$ is a positive parameter).}

\begin{table}
	    \caption{\changed{Parameters of the simulation model}}
        \label{tab:parameters}
        \begin{center}     
\changed{    \begin{tabular}{|c|c||c|c|}
        \hline
        $c_1$ & \SI{0.513}{\sqrt{cm} \per s} &
        $c_2$ & \SI{0.513}{\sqrt{cm} \per s} \\ \hline
        $c_3$ & \SI{0.299}{cm \per V s} &
        $c_4$ & \SI{7}{cm} \\ \hline
        $c_5$ & \SI{2}{cm} &
        $c_6$ & \SI{0.008}{Hz} \\
        \hline
\end{tabular}
}
\end{center}
\end{table}

\changed{Fig.~\ref{fig:tank_sim} compares the tracking performance obtained with the proposed controller to that of a standard I-controller 
\begin{equation}
        q = \qrefi(t) + \ki v, \qquad
        \dot v = -\beta (z_2 - \zrefi{2}(t))
    \end{equation}
    with constant positive parameters $\ki$ and $\beta$.
    For the simulation, the nonlinear plant is used with constant disturbance $w = \SI{0.5}{cm \per s}$, saturation limit $Q = \SI{8}{V}$ and initial conditions $\z(0) = \bm{0}$, $v(0) = 0$.
    For comparison purposes, the constants $\alpha = 0.12$, $\beta = 0.0062$ are chosen to obtain a similar settling time (to within $\SI{2}{\percent}$ of the reference) with both controllers for $\ki = 1$.
One can see that, with increasing $\ki$, performance improves with the proposed controller, whereas the I-controller tends to oscillations.
    As a result, the proposed controller achieves superior tracking, and its performance is insensitive even to fairly large gains $\ki$.
    For $\ki = 10$, the proposed controller is moreover simulated with and without anti-windup for comparison, demonstrating also the practical usefulness of the proposed anti-windup strategy in the form of a reduced overshoot.
}

\begin{figure}[tbp]
    \centering
    \includegraphics{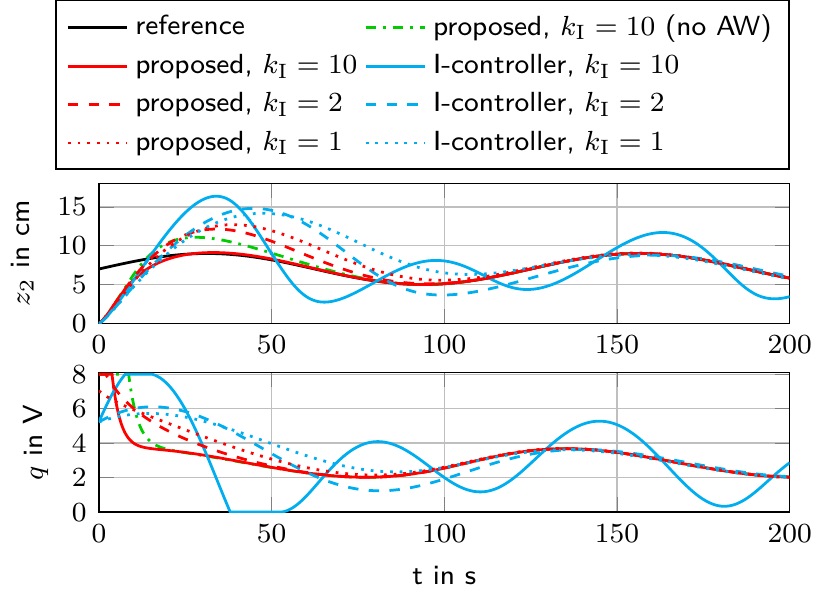}
    \caption{\changed{Simulated liquid level $z_2$ of the second tank along with its reference, and pump voltage $q$.
    Results obtained with the proposed controller \eqref{eq:sim:ctrl} and with a standard I-controller are shown, for different values of the integral gain $\ki$ and in one case with disabled anti-windup (AW).}}
\label{fig:tank_sim}
\end{figure}

\section{Conclusion and Outlook}
\label{sec:conclusion}

An approach for adding integral action to a given state-feedback controller for a linear, time-varying, multivariable plant was proposed.
With proper initialization of the integrator, performance of the nominal state feedback is preserved in the unperturbed case and, asymptotically, also with increasing integrator gain.
Additionally, in the time-invariant case, all nominal closed-loop eigenvalues are preserved.

Conditions were derived that allow to guarantee stability for any positive definite integrator gain;
specifically, bounded-input bounded-state stability for any disturbance as well as asymptotic stability for perturbations, whose action on the plant is constant with respect to the control input, was shown.
To facilitate the practical implementation of the controller, tuning guidelines and a scheme to mitigate windup were discussed.

Future work may focus on further investigating the case of using only the integral of a given output to construct the controller, i.e., the design of an output-feedback integral action.
As shown, this case requires to invert the dual of the nominal closed loop in such a way that the integral of its output is positive definite.
Achieving this in the general case, without relying on flatness or related properties, would be an interesting problem to be studied.
Furthermore, the use of estimated rather than directly measured plant states and the corresponding disturbance rejection properties may also be investigated.

\appendix

\endgroup

\bibliographystyle{elsarticle-harv}
\bibliography{literature}           

\end{document}